\documentclass[twocolumn,prc,floatfix,showpacs,preprintnumbers,nofootinbib,%
superscriptaddress]{revtex4}
\usepackage{mathrsfs}
\usepackage{amssymb}
\usepackage{amsmath}
\usepackage{graphicx}

\usepackage{xcolor}
\definecolor{lcolor}{rgb}{0.5,0,0}
\definecolor{citcolor}{rgb}{0,0.3,0.0}

\usepackage[breaklinks,colorlinks,urlcolor=blue,citecolor=citcolor,linkcolor=lcolor]{hyperref}

\newcommand{\xt}{{\mathbf{x}_T}}
\newcommand{\bt}{{\mathbf{b}_T}}

\newcommand{\yt}{{\mathbf{y}_T}}

\newcommand{\pt}{{\mathbf{p}_T}}

\newcommand{\ptt}{p_T} 
\newcommand{\qtt}{q_T} 

\newcommand{\ud}{\, \mathrm{d}}

\newcommand{\tr}{\, \mathrm{Tr} \, }

\newcommand{\nc}{{N_\mathrm{c}}}

\newcommand{\nr}[1]{(\ref{#1})}

\newcommand{\rf}{\mathrm{ref}}

\newcommand{\gev}{\ \textrm{GeV}}

\newcommand{\qs}{Q_\mathrm{s}}

\newcommand{\figs}{Figs.~}
\newcommand{\eq}{Eq.~}

\begin{document}

\author{T. Lappi}
\affiliation{
Department of Physics, %
 P.O. Box 35, 40014 University of Jyv\"askyl\"a, Finland
}
\affiliation{
Helsinki Institute of Physics, P.O. Box 64, 00014 University of Helsinki,
Finland
}

\title{
Azimuthal harmonics of color fields in a high energy nucleus
}


\preprint{}

\begin{abstract}
Recent experimental results have revealed a surprisingly rich structure of 
multiparticle azimuthal correlations in high energy proton-nucleus collisions. 
Final state collective effects can be responsible for many of the observed 
effects, but it has recently been argued that 
a part of these correlations are present already
in the wavefunctions of the colliding particles. 
We evaluate the  momentum space 2-particle 
cumulant azimuthal anisotropy coefficients $v_n\{2\}, n=2,3,4$ from fundamental 
representation Wilson line
distributions describing the high energy nucleus. These would
correspond to the flow coefficients in very forward proton-nucleus scattering.
We find significant differences beteen Wilson lines from 
the MV model and from JIMWLK evolution.
The magnitude and transverse momentum dependence of the  $v_n\{2\}$ values 
suggest that the fluctuations present in the initial fields
are a significant contribution to the observed anisotropies.
\end{abstract}

\maketitle

\section{Introduction}

One of the most surprising results from the LHC proton-nucleus collision experiments 
have been the kind of  azimuthal multiparticle correlation 
structures~\cite{Abelev:2012ola,Abelev:2014mda,CMS:2012qk,Chatrchyan:2013nka,Aad:2012gla,Aad:2013fja,Aad:2014lta} 
(see also RHIC results from deuteron-gold collisions \cite{Adare:2013piz,Adare:2014keg})
that have, in larger collision systems, been attibuted to hydrodynamical flow.
The particle multiplicities in these collisions systems are large enough
for some collective effects to take place.
Many of these structures have indeed been successfully descibed by 
hydrodynamical calculations~\cite{Bozek:2012gr,Werner:2013ipa}.
This agreement requires, however, a very specific model of 
the geometry of the initial state~\cite{Schenke:2014zha}. It is also not clear whether
these small systems are within the regime of validity of a hydrodynamical description
with realistic values of the energy density, viscosity and system size~\cite{Niemi:2014wta}.

The primary collisions leading to energy deposition in the central rapidity region are,
at the high energies reached at the LHC, characterized by very strong nonlinear
color fields~\cite{Gelis:2010nm}.
These fields are, to leading order in the coupling constant, boost invariant. This 
immediately leads to the presence of long range azimuthal correlations in particle
production~\cite{Dumitru:2008wn,Dusling:2009ar,Gelis:2009wh,Dusling:2009ni,Dumitru:2010iy,Kovner:2010xk,Kovner:2011pe,Dusling:2012wy}. In larger collision systems, the structure of these correlations in 
azimuthal angle and transverse momentum is strongly influenced by collective
behavior in the later evolution stages of the system.
However, in smaller systems, such as proton-nucleus collisions, these collective effects are
presumably less significant than in nucleus-nucleus collisions. 
This raises the intriguing possibility 
that in proton-nucleus collisions
also the azimuthal structure of the initial stage color fluctuations could 
be directly visible in the measurable particle spectrum.

We do not yet have a very solid quantitative understanding of the relative
importance of initial color field and later evolution effects for generating
anisotropies in particle production.
A complete calculation of azimuthal anisotropies in this context requires complicated
modeling that includes the color field and nucleonic scale
fluctuations in the nucleus~\cite{Schenke:2012wb}
and in the proton~\cite{Schlichting:2014ipa},
 combined with a calculation of the time evolution of the
initial color fields and eventual matching to a hydrodynamical description~\cite{Gale:2012rq}.
We will not attempt to carry out this whole program here, but concentrate in this
paper only on a part of it, namely the anisotropies produced when a bunch of
valence quark-like particles in the fundamental representation of the gauge group
scatter off the color field of a large nucleus. The physical picture
(see~\cite{Kovner:2010xk,Kovner:2011pe} and more
recently~\cite{Dumitru:2014dra,Dumitru:2014yza,Dumitru:2014vka,Skokov:2014tka})
in our calculation is that of valence quarks from the probe 
deflected in a preferred transverse
direction by a domain in the target color field. This generates a multiparticle
correlation that probes the spatial fluctuations of the target. Our calculation 
extends the work in \cite{Dumitru:2014vka,Skokov:2014tka} in two significant ways.
Firstly, we perform the Fourier-transform from coordinate to momentum space,
in order to get an azimuthal harmonic coefficient corresponding to real produced
particles. Secondly, unlike \cite{Dumitru:2014vka,Skokov:2014tka},
we correlate the particles in a given $\ptt$-bin with a reaction plane determined
by all the produced particles using the 2-particle cumulant method.

\section{Azimuthal correlations in CGC fields}

In the ``hybrid formalism'' for particle production in the dilute-dense 
limit~\cite{Dumitru:2002qt,Dumitru:2005gt,Altinoluk:2011qy,Chirilli:2011km} the 
quark spectrum produced in a collision is proportional to the Fourier-transform 
of the two point function of Wilson lines in the color field of the target
\begin{equation}\label{eq:sinc}
 \frac{\ud N}{\ud^2\pt}\propto 
\int\limits_{\xt,\yt}e^{- \pt \cdot(\xt-\yt)} 
\frac{1}{\nc} \tr V_\xt^\dag V_\yt.
\end{equation}
The Wilson lines $V(\xt)$ in \nr{eq:sinc} are, in the CGC description, stochastic
random SU(3) matrices in the representation of the projectile.
 To calculate the single inclusive cross section
one has to average \eq\nr{eq:sinc} by the appropriate probability distribution 
of Wilson lines.

In the leading order CGC treatment that we use here, multiparticle correlations
can be calculated from the higher order moments of the Wilson line operators in 
\eq\nr{eq:sinc}. This corresponds to the so called 
``glasma graphs''~\cite{Dusling:2009ni,Dumitru:2010iy,Dusling:2012wy}, which 
have a very clear interpretation in the hybrid formalism. The target 
nucleus is represented by a sheet of color magnetic and color electric fields,
which have a domain structure with a characteristic length scale $1/\qs$ in 
the transverse plane. When a small enough probe (comparable in size to the 
domain size) hits this target, the resulting particle production is 
has a preferred direction given by the direction of the color field in the domain.
Since this direction fluctuates from event to event, there is of course no anisotropy 
on average, but the existence of a preferred direction in individual events shows 
up in a global angular correlation among all of the produced particles,
similarly to hydrodynamical flow. 
We are neglecting here
 ``connected'' or ``BFKL''-like correlations~\cite{Dusling:2012wy}, that give rise to a 
back-to-back peak in the two-particle correlation. These correlations are 
typical  ``nonflow'' correlations that involve only a few particles, which 
the experimental analyses of azimuthal anisotropy
try to exclude. We will not discuss them further here, see however 
Refs.~\cite{Dusling:2012wy,Ducloue:2013hia} for more studies on these lines.

It is evident from the above discussion 
that we expect the correlation to be very sensitive to the transverse
size of the probe. In the case of calculating the initial condition 
for an ion-ion collision the probe is large, with the consequence that 
 the correlation is washed away by the sum over many independent 
domains in the transverse plane. Thus, in contrast to the 
correlations generated by collective flow, the effect discussed
here is stronger in small collision systems than in large ones.

The purpose of this paper is to analyze the azimuthal correlation structure
of particle production using \eq\nr{eq:sinc} in more detail. In particular, we
want to study its dependence on the harmonic $n$, transverse momentum, and the 
transverse size of the probe.  The practical procedure used here is the 
following. We first divide the $\ptt$ range accessible on the lattice into bins.
We use here 50 bins, but we have checked that the results are independent of the size
of the bin.
We then define the Fourier coefficient of the single particle spectrum as
\begin{multline}\label{eq:defbn}
 b_n(\ptt) \equiv 
\!\!
\int\limits_{|\pt| \in \textrm{bin}} 
\ud^2 \pt
e^{i n \varphi_\pt}
\int\limits_{\xt,\yt}
e^{- i \pt \cdot(\xt-\yt)} 
\\
\times S_p(\xt-\bt)
S_p(\yt-\bt)
\frac{1}{\nc} 
\tr V_\xt^\dag V_\yt.
\end{multline}
The transverse coordinate profile of the probe 
has been taken  as a Gaussian
\begin{equation}
 S_p(\xt-\bt) = \exp \left\{ \frac{-(\xt-\bt)^2}{2B}\right\}
\end{equation}
around an impact parameter $\bt$ chosen randomly in the transverse plane
of the target. We will present results for different values of the
 parameter $B$ characterizing the size of the probe.
Note that the coefficients \nr{eq:defbn} need not be normalized, since we will eventually
divide by the angular average spectrum  $b_0$  to construct the Fourier harmonic
coefficient. We want to calculate the angular correlations with respect to a
an event plane defined by all the all the produced particles, which form 
the ``reference'' that we correlate individual particles with. This is done
following the procedure used in the experimental analysis (see e.g.
the 2-particle cumulant method in \cite{Chatrchyan:2013nka}).
For this we need to calculate also the reference coefficients
\begin{multline}
 b_n(\rf) \equiv 
\int 
\ud^2 \pt
e^{i n \varphi_\pt}
\int\limits_{\xt,\yt}e^{- i\pt \cdot(\xt-\yt)} 
\\
\times S_p(\xt-\bt)
S_p(\yt-\bt)
\frac{1}{\nc} 
\tr V_\xt^\dag V_\yt
\end{multline}
integrated over all momenta.

The target Wilson lines are drawn from a completely homogenous
and isotropic distribution that fills the whole transverse lattice with
periodic boundary conditions, and the probe is azimuthally symmetric.
Thus there is no geometrical (i.e. originating
in the shape of the probe or the target) origin for azimuthal anisotropy
present in the calculation.
Since the probability distribution of Wilson lines is azimuthally symmetric
(although the individual configurations are not), the correlations
among the coefficients $b_n$ are diagonal:
\begin{equation}
\left\langle b_n^*(\ptt) b_m(\qtt) \right\rangle \propto \delta_{m,n}, 
\end{equation}
where $\langle \rangle$ denotes averaging over the configurations of Wilson 
lines in the target.
Note that the single particle spectrum \eq\nr{eq:sinc} 
is explicitly real, configuration by configuration, leading to
 $b_n = b_{-n}^*$. This can be shown by
taking the complex conjugate of \eq\nr{eq:sinc} and exchanging the 
integration variables $\xt$ and $\yt$. 
The two particle pair correlation function is now 
\begin{equation}\label{eq:2part}
 \frac{\ud N_\textnormal{pair}}{\ud \Delta \varphi}
\propto 
\sum_{n=-\infty}^\infty \left\langle b_n^*(\ptt) b_n(\qtt) \right\rangle
 \cos ( n \Delta \varphi).
\end{equation}
From this we can identify the correlation function
Fourier coefficients (using the notation 
of \cite{Chatrchyan:2013nka}) 
\begin{equation}
 V_{n\Delta}(\ptt,\qtt) = 
\frac{\left\langle b_n^*(\ptt) b_n(\qtt) \right\rangle}{
\left\langle b_0^*(\ptt) b_0(\qtt) \right\rangle
},
\end{equation}
and define the 2-particle cumulant azimuthal harmonic as in  \cite{Chatrchyan:2013nka}
as
\begin{eqnarray}
v_n\{2\} &=& \frac{V_{n\Delta}(\ptt,\rf)}{\sqrt{V_{n\Delta}(\rf,\rf)}}
\\
\label{eq:vn2}
&=& \frac{
\frac{\left\langle b_n^*(\ptt) b_n(\rf) \right\rangle}{
\left\langle b_0^*(\ptt) b_0(\rf) \right\rangle
}
}{
\sqrt{\frac{\left\langle b_n^*(\rf) b_n(\rf) \right\rangle}{
\left\langle b_0^*(\rf) b_0(\rf) \right\rangle
}}
}.
\end{eqnarray}
A nice interpretation of \eq\nr{eq:vn2} can be obtained by writing
it as a product of three terms, 
\begin{equation}\label{eq:3term}
 v_n\{2\} = v_n\{\textrm{bp}\} \frac{R_n(\ptt,\rf)}{R_0(\ptt,\rf)}.
\end{equation}
Here we denote by
\begin{equation}\label{eq:defvnbp}
v_n\{\textrm{bp}\}^2
=
\frac{
\left\langle b_n^*(\ptt) b_n(\ptt) \right\rangle
}{
\left\langle b_0^*(\ptt) b_0(\ptt) \right\rangle
}
\end{equation}
the flow coefficient for particles in the $\ptt$ bin 
with respect to the event plane of that $\ptt$ bin (``bp'' stands for ``bin plane''). 
This is the equivalent (although here in momentum, not position space)
of the quantity calculated in \cite{Dumitru:2014vka}.
This is  then corrected by two ``correlation coefficients''. 
The first one  is the correlation coefficient between the reference reaction plane and the $\ptt$-bin reaction plane:
\begin{equation} 
R_n(\ptt,\rf) \equiv
\frac{\left\langle b_n^*(\ptt) b_n(\rf) \right\rangle}{
\sqrt{
\left\langle b_n^*(\ptt) b_n(\ptt) \right\rangle
\left\langle b_n^*(\rf) b_n(\rf) \right\rangle
}}
\leq 1,
\end{equation}
where the inequality follows from the Schwartz inequality. 
The interpretation of this correction is clear: for a fixed anisotropy with respect to
the $\ptt$-bin reaction plane, a decorrelation of the $\ptt$-bin reaction plane from
the reference reaction plane decreases  the flow coefficient 
$v_n\{2\}$.
The other correlation coefficient factor in \nr{eq:3term}
\begin{equation}
R_0(\ptt,\rf) \equiv
\frac{\left\langle b_0^*(\ptt) b_0(\rf) \right\rangle}
{\sqrt{
\left\langle b_0^*(\ptt) b_0(\ptt) \right\rangle
\left\langle b_0^*(\rf) b_0(\rf) \right\rangle
}}
\leq 1
\end{equation}
is related to the multiplicity and appears in the denominator, increasing $v_n\{2\}$.
This can be understood as follows: with larger fluctuations in the $\ptt$-bin 
multiplicity that are independent of the reference multiplicity, a fixed 
correlation between $b_n(\ptt)$ and $b_n(\rf)$ implies a larger
correlation between $\ptt$-bin and reference  reaction planes.
In other words, since $b_n \sim v_n b_0$, for a given correlation between ($\rf$ and 
$\ptt$) $b_n$'s, the smaller the correlation between $b_0$'s, the larger must the correlation 
between $v_n$:s be.

We take the Wilson lines $V(\xt)$ appearing
in \eq\nr{eq:sinc} either from the MV model or resulting from JIMWLK evolution 
of the distribution of Wilson lines. Both are discretized on a $1024^2$ transverse
lattice. For the MV model we use a (fundamental representation) 
saturation scale  of $\qs a = 0.119$, where $a$ is the lattice spacing. 
The JIMWLK calculation starts with a MV model at $\qs a= 0.0220$ and, after 
$y=10$ units of evolution in rapidity (with running coupling) ends up with 
$\qs a = 0.117$. The MV model Wilson lines are constructed following the procedure
decribed in more detail in \cite{Lappi:2007ku} and the running coupling 
JIMWLK evolution performed using the algorithm of \cite{Lappi:2012vw}. The parameter
values used here are exactly the same as for the  $1024^2$-lattice in 
\cite{Dumitru:2014nka}. Note that we are only averaging two-point functions of
the coefficients $b_n$, not ratios $\sim b_n/b_0$. This
makes the averaging procedure numerically quite stable and is physically
the correct thing to do, since the pair correlation function 
\eq\nr{eq:2part} is the correct inclusive observable to be obtained via 
the target average \cite{Gelis:2008ad,Gelis:2009wh}.

\begin{figure*}[pt!]
\centerline{\includegraphics[width=0.45\textwidth]{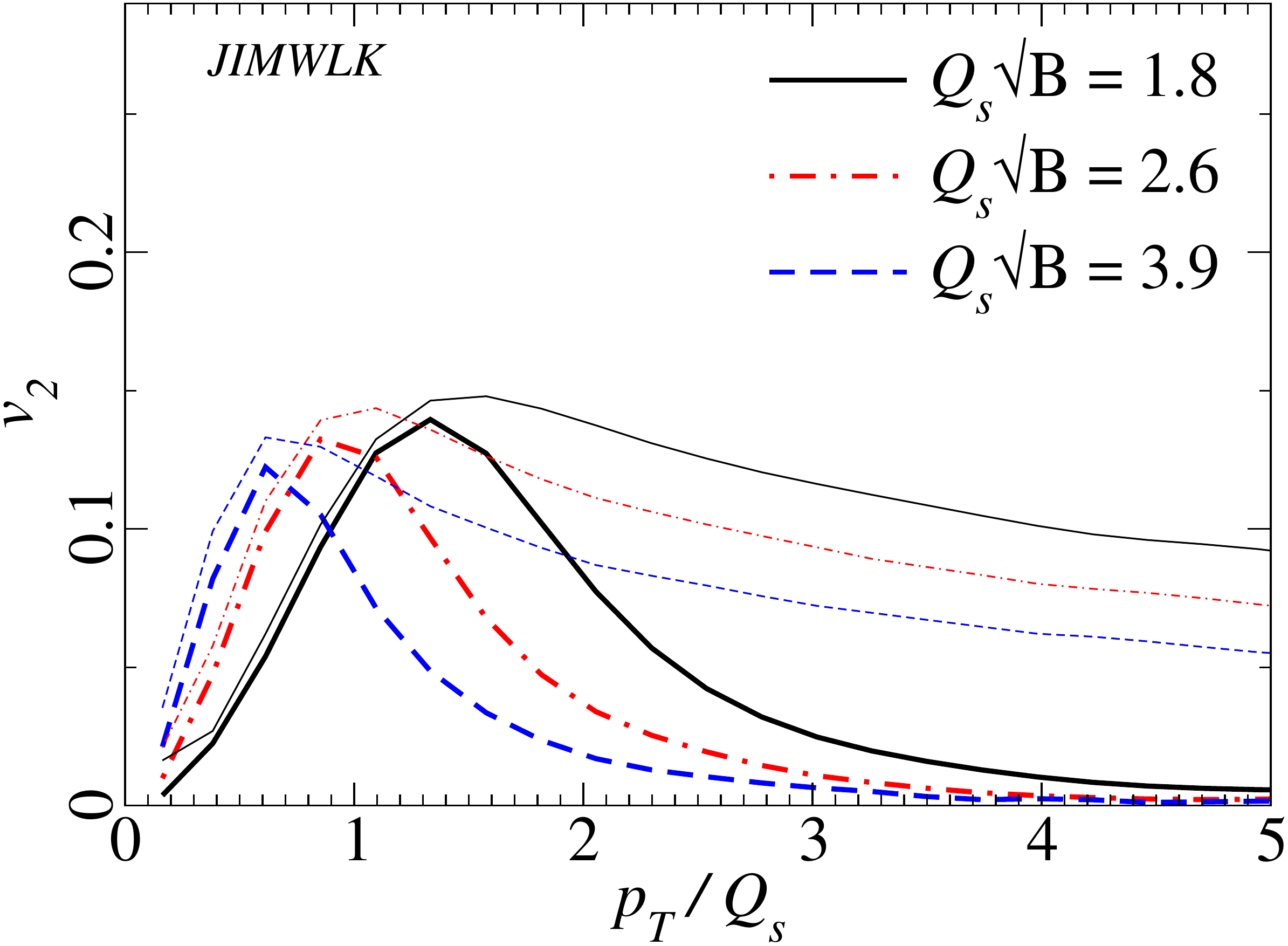}
\hfill
\includegraphics[width=0.45\textwidth]{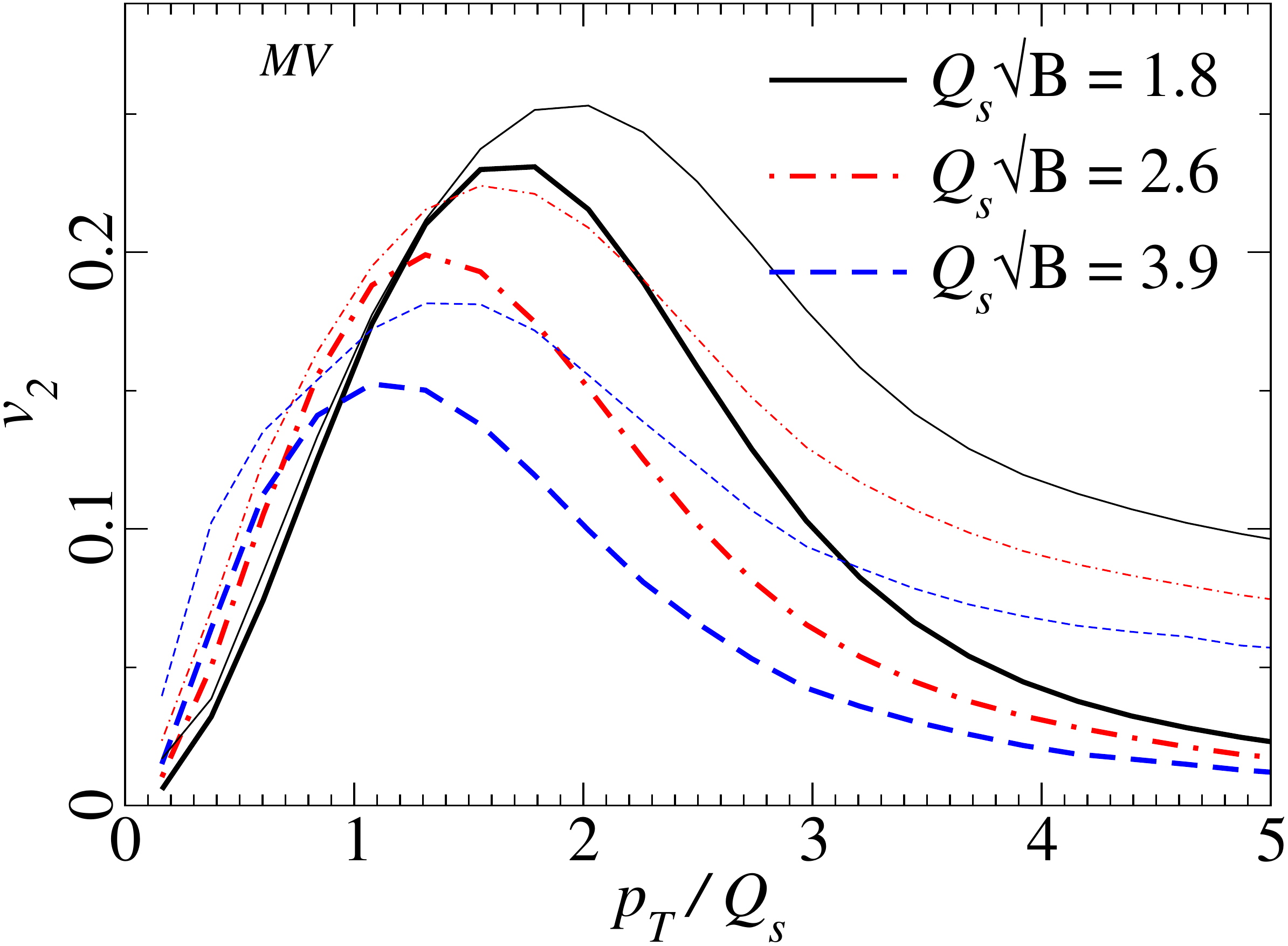}}
\caption{
Second harmonic coefficient $v_2\{2\}$ calculated with JIMWLK-evolved (left) 
and MV-model (right) Wilson line configurations. The thin lines represent the 
cofficients $v_2\{\textrm{bp}\}$ (see \eq\nr{eq:defvnbp}) 
calculated with respect to the event plane
in the $\ptt$ bin only.
\label{fig:v2}
}
\end{figure*}

\begin{figure*}[pt!]
\centerline{\includegraphics[width=0.45\textwidth]{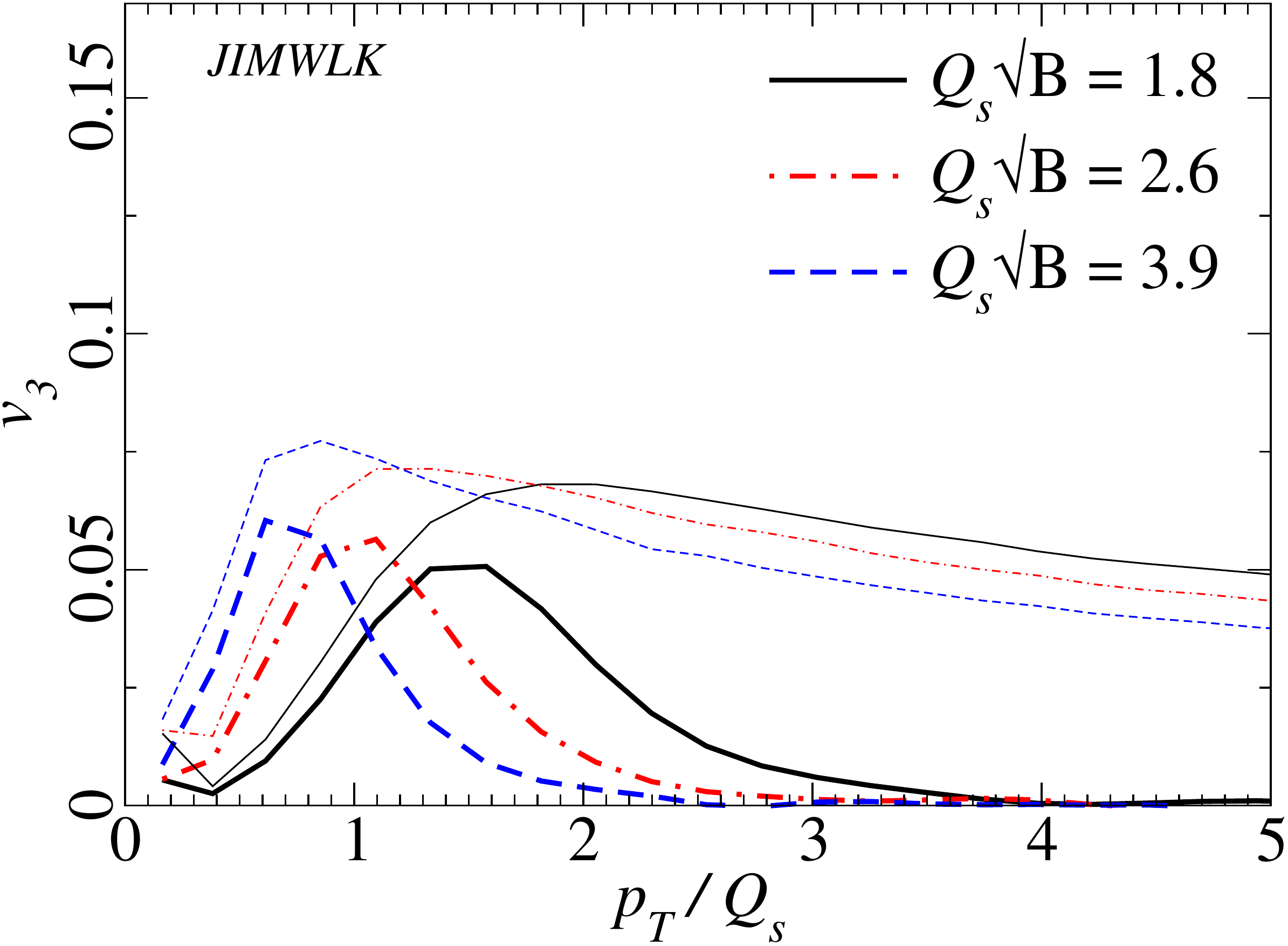}
\hfill
\includegraphics[width=0.45\textwidth]{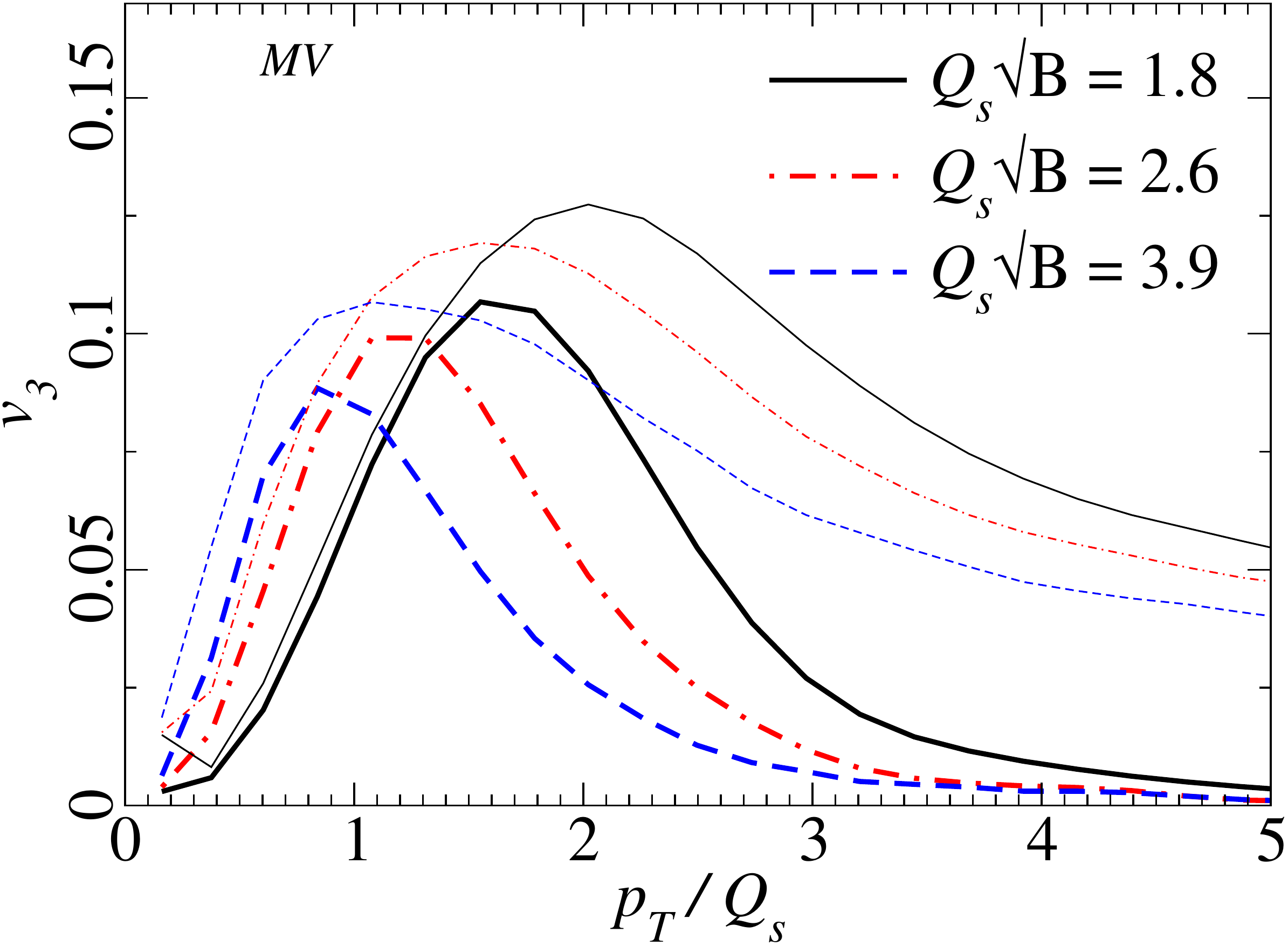}}
\caption{
Third  harmonic coefficient $v_3\{2\}$ calculated with JIMWLK-evolved (left) 
and MV-model (right) Wilson line configurations. The thin lines represent the 
cofficients $v_3\{\textrm{bp}\}$ (see \eq\nr{eq:defvnbp}) 
calculated with respect to the event plane
in the $\ptt$ bin only.
\label{fig:v3}
}
\end{figure*}

\begin{figure*}[pt!]
\centerline{\includegraphics[width=0.45\textwidth]{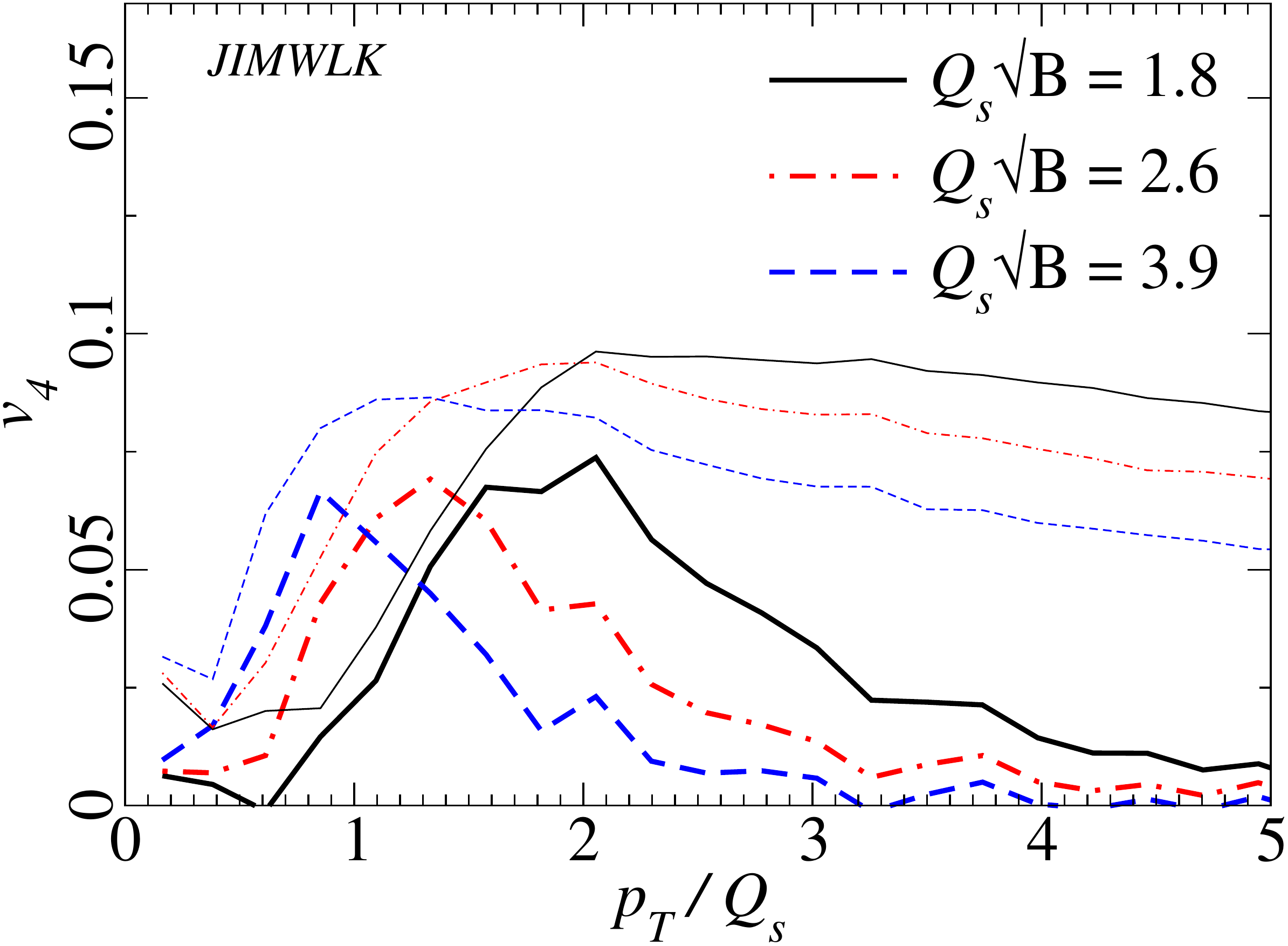}
\hfill 
\includegraphics[width=0.45\textwidth]{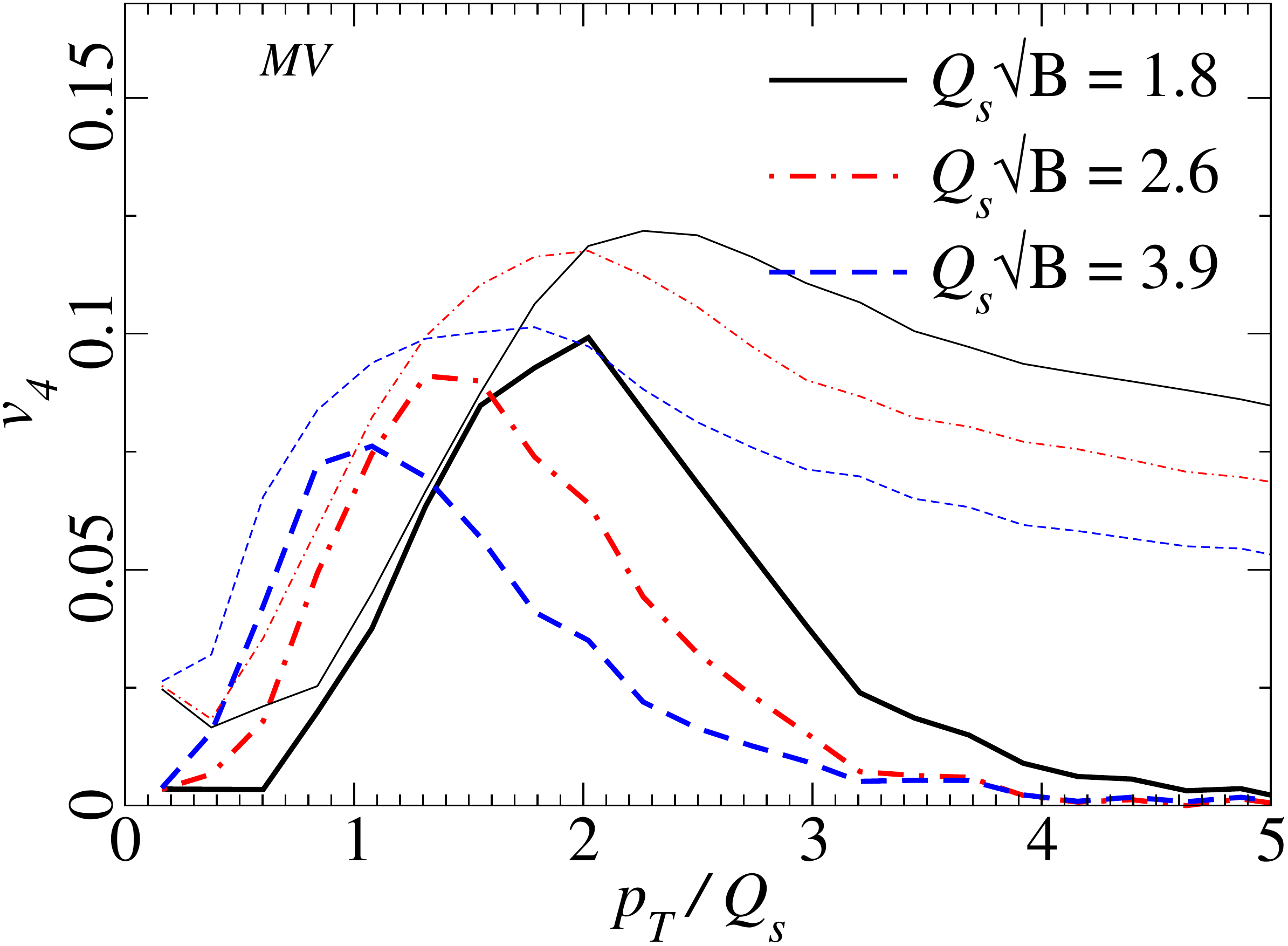}}
\caption{
Fourth harmonic coefficient $v_4\{2\}$ calculated with JIMWLK-evolved (left) 
and MV-model (right) Wilson line configurations. The thin lines represent the 
cofficients $v_4\{\textrm{bp}\}$ (see \eq\nr{eq:defvnbp}) 
calculated with respect to the event plane
in the $\ptt$ bin only.
\label{fig:v4}
}
\end{figure*}

\section{Results and discussion}

The numerical evaluations of the first four anisotropy coefficients are shown in 
\figs\ref{fig:v2}, \ref{fig:v3} and \ref{fig:v4}. The results are presented in 
scaling units, as $v_n$ plotted against $\ptt/\qs$ for different size probes,
i.e. different $\sqrt{B}\qs$.
Also shown are the ``bin plane'' coefficients $v_n\{\textrm{bp}\}$, defined 
by \eq\nr{eq:defvnbp}.
To set the scale of the parameters in physical units we note 
that the fundamental representation saturation scale around midrapidity at the 
LHC should be~\cite{Lappi:2011gu} around $\qs\sim 1\gev$ and the typical size of
a proton in hard particle production at small $x$ around 
$B\approx 4 \gev^{-2}$~\cite{Kowalski:2006hc}. 
Thus a realistic probe size for LHC pA collisions would very roughly be 
$\sqrt{B}\qs \approx 2$.

The first immediate observation from the numerical results is that the color
 field fluctuations indeed generate anisotropies that are large, of the order
of the experimentally measured anisotropy coefficients. It seems therefore plausible
that the color field fluctuations do play a sizeable role in the observed anisoptropy
in small systems, and must be taken into account together with the flow contribution.
Also the momentum distribution has the same structure as the observed transverse 
momentum dependence of the flow, first rising until  $\sim \qs$ and then decreasing.
The ``bin plane'' coefficients $v_n\{\textrm{bp}\}$ do not decrease nearly as
fast at high momentum, from which one can deduce that
the decrease of the anisotropy coefficients at large $\ptt$ follows 
from the decorrelation of the event plane in the $\ptt$ bin from the reference. 
This explains why this decrease was not seen in \cite{Dumitru:2014vka}, where
this decorrelation was not taken into account. The MV model has a gluon spectrum
that is more sharply peaked around $\qs$, i.e. a narrower distribution 
of different size color field domains. This shows up in  significantly larger values
for the $v_n$ coefficients. The main effect of JIMLK evolution is to add
more small color field domains (larger $\ptt$ gluons), which decrease the 
anisotropy of the particle spectrum.

There is, however, an important 
caveat concerning any direct comparison of these results to experimental values. 
Namely, we were considering, in \eq\nr{eq:sinc}, only incoming quarks.
 For antiquarks one must replace the Wilson line by its Hermitian conjugate, which 
changes the sign of $b_n$ for odd $n$. Away from the very forward valence
region in the probe, there are an approximately equal amount of quarks and antiquarks
present,  with contributions to $v_3$ that  therefore cancel.  
Gluons do not have nonzero odd harmonics in this mechanism, because
the adjoint representation is real and thus odd $b_n$'s vanish.
Any odd harmonic
surviving in the final state around midrapidity must therefore have
an origin that is different from the one discussed here.

The other word of caution in interpreting these results is related to the 
dependence on the size of the probe, parametrized here by the width of the Gaussian 
$\sqrt{B}$. As anticipated, the magnitude of the correlation, and its dependence on the 
transverse momentum, depends strongly on the size of the interaction region. 
Although one can quite well estimate this, it depends on nonperturbative
physics in the proton and cannot ultimately be controlled in a weak coupling calculation.


Results for  azimuthal correlations
in a full CYM simulation have also been presented recently
by Schenke, Schlichting and Venugopalan~\cite{Schlichting:is2014,Schenke:2015aqa}.
Their calculation  includes effects of both color field
and nucleonic fluctuations in
the probe proton and the target nucleus. 
The probe and target  geometries also 
have a significant effect through the CYM
pre-equilibrium version of the usual hydrodynamical mechanism that converts spatial
anisotropy to momentum space, leading to also odd harmonics.
These geometrical effects have not been included in our work,
which should therefore not be compared directly with experimental
data. Our focus here has been, in stead, on quantifying the generic observation that
fluctuating color fields result in azimuthal anisotropies in multiparticle
correlations, even in the absence of anisotropies in the impact
parameter dependence.

As a conclusion, we have here studied the momentum space azimuthal anisotropy structure of the 
``color glass'' gluon fields in a high energy nucleus, as they are seen by a small probe 
consisting of valence-like quarks.
We also  quantified here the effect of correlating the particles with the event plane
determined by all the  produced particles, using the two-particle cumulant method at
 the parton level. 
The quantitative results strongly depend on the details of the $\ptt$-distribution of gluons 
in the CGC wavefunction and on the transverse size of the probe. However, all the results show
large contributions to the harmonics from these purely initial state effects. For odd harmonics
they largely cancel between quarks and antiquarks, but for even harmonics these are sizeable 
effects that need to be considered when interpreting the experimental results from 
proton-nucleus collisions.

\section*{Acknowledgements}
The author wishes to thank A.~Dumitru for discussions, K.~J.~Eskola for comments on the
manuscript and  Baruch College, CUNY for hospitality in the initial stages of this work.
The author is supported by the Academy of Finland, projects 
267321 and 273464. This work was done using computing resources from
CSC -- IT Center for Science in Espoo, Finland.

\bibliography{spires}
\bibliographystyle{JHEP-2mod}

\end{document}